\documentclass[useAMS,usenatbib,usegraphicx]{mn2e}

\title[Photometry and spectroscopy of BZ UMa during outburst]
{Simultaneous photometry and echelle-spectroscopy of the dwarf nova
 BZ Ursae Majoris during the 2005 January Outburst}

\author[V.\,V.\,Neustroev et al.]{V.\,V.\,Neustroev$^{1}$\thanks{E-mail:
benj@it.nuigalway.ie},
S.\,Zharikov$^2$ and R.\,Michel$^2$\\
$^{1}$Computational Astrophysics Laboratory, National University
of Ireland, Galway, Newcastle Rd., Galway, Ireland\\
$^{2}$Observatorio Astronomico Nacional, Instituto de Astronomia, UNAM,
Ensenada, BC, Mexico}
\begin{document}

\date{Accepted ???. Received ???; in original form 2005 September 12}

\pagerange{\pageref{firstpage}--\pageref{lastpage}} \pubyear{2005}

\maketitle

\label{firstpage}

\begin{abstract}

 We report simultaneous photometric and echelle-spectroscopic observations of the
 dwarf nova BZ UMa during which we were lucky to catch the system at the onset of
 an outburst, the development of which we traced in detail from quiescence to early decline.
 The outburst had a precursor, and was of a short duration ($\sim$5 days) with a highly
 asymmetrical light curve. On the rise we observed a `jump' during which the brightness
 almost doubled over the course of half an hour.
 Power spectra analysis revealed well-defined oscillations
 with period of $\sim$42 minutes. Using Doppler tomography we found that the
 unusual emission distribution detected in quiescence held during the outburst.
 After the maximum a new emission source arose, from the inner hemisphere of the secondary 
 star, which became the brightest at that time. 
 We analyse this outburst in terms of `inside-out' and
 `outside-in' types, in order to determine which of these types occured in BZ UMa.
\end{abstract}


\section{Introduction}

   Cataclysmic Variables (CVs) are close interacting binaries that contain
   a white dwarf (WD) accreting material from a companion, usually a
   late main-sequence star (see review by \citet{Warner}).
   Dwarf novae (DNs) are an important subset of CVs, which
   undergo recurrent outbursts of 2--6 mag on timescales of days to years,
   identified with the release of gravitational energy when rapid accretion
   occurs from an accretion disc onto the WD. It is now widely accepted
   that the reason for this is a thermal instability in the accretion disc,
   which switches the disc from a low-viscosity to a high-viscosity regime
   \citep{Smak, Osaki, Lasota}.

   In order to explain the rich variety in outburst behaviour of DNs, the current disc
   instability model (DIM) has become rather complex. To restrict its free
   parameters, whose number has recently significantly increased, observations
   of DNs just prior
   to and during the early phases of the outburst are necessary. Unfortunately,
   it is difficult to obtain such observations because of the nonperiodic nature of the
   outbursts and the impossibility of predicting subsequent events.

   In this paper we describe the observations of the little-studied
   dwarf nova BZ\,UMa -- a quite unusual system with an orbital period of 97.9 minutes
   \citep{Ringwald,Jurcevic}. A SU UMa classification of this star based on the short
   orbital period is uncertain because no superoutbursts/superhumps have ever been
   detected. We observed BZ\,UMa in January 2005 in order to investigate its extremely
   unusual emission structure recently discovered by us \citep{Neustroev1} when we found
   the system was going to the outburst. This is especially important as BZ\,UMa is
   characterized by infrequent outbursts with mean intervals of 313 days between them
   \citep{Price}, and no one has observed the early phases of its outbursts to date.
   We have traced the development of the outburst in detail and here present our results.

\section{Observations and Data reduction}

   Photometric observations of BZ\,UMa in the Johnson $V$ band were performed
   at the Observatorio Astronomico Nacional (OAN SPM) in Mexico on the 1.5~m telescope
   for a total of 5 nights (2005 January 12--13 and 15--17). The exposure times were
   ranged from 80~s on Jan 12 to 5~s on Jan 17. The magnitudes of the object were
   determined using the calibration stars reported by \citet{Misselt}.

    Echelle observations have been obtained with the REOSC Espresso spectrograph
    \citep{Echelle} on the 2.1 m telescope at the same site. This instrument gives a
    resolution of 0.234 \AA\ pixel$^{-1}$ at H$\alpha$ using the UCL camera and a CCD-Tek
    chip of 1024x1024 pixels with a 24~$\mu m^2$ pixels size. The spectra cover 27 orders
    and span the spectral range 3720-6900 \AA.
    Test observations of BZ\,UMa performed on 2004 December 10 in order to investigate
    the quality of the resultant spectra showed that observations with this spectrograph
    are acceptable and suitable for analysis.
    In the next program observations, in order to improve the signal-to-noise ratio we
    obtained a series of phase-locked spectra: 15 spectra were taken at equal phase
    intervals over a single orbital period P$_{orb}$(=97.9 min) with an exposure
    time of 325 sec per spectrum. This sequence of spectra was repeated at exactly
    the same phase intervals for subsequent periods and subsequent
    nights. This allows us to calculate the phase-averaged spectra, summarizing
    the spectra of the same orbital phase obtained during one night and the whole set of
    observations without further decreasing the time resolution. Such averaging has
    been performed at the stage of the primary reduction, \textit{before} the extraction
    of the spectral information from the CCD images. The bulk of the spectra was obtained
    during the 3 consecutive nights of 2005 January 15-17, simultaneously with photometry.

    Log of observations is presented in Table \ref{tab1}.

\begin{table}
\begin{center}
\caption{Log of observations of BZ UMa.}
\begin{tabular}{llllccc}
\hline
UT Date     & HJD Start & Exposure & No. of & Duration\\
            &(+2453000) & Time (s) & Exps   &         \\
\hline
 Photometry &         & \\
2005-Jan-12 & 382.889 &     80     & 169    & 4\fh14  \\
2005-Jan-13 & 383.841 &     60     & 281    & 5\fh33  \\
2005-Jan-15 & 385.806 &     30     & 556    & 5\fh97  \\
2005-Jan-16 & 386.727 &     30, 15 & 978    & 8\fh38  \\
2005-Jan-17 & 387.743 &     05, 10  & 2030   & 8\fh14  \\
Spectroscopy&         & \\
2004-Dec-10 & 349.970 &     420    & 12     & 1\fh65  \\
2005-Jan-15 & 385.787 &     325    & 60     & 6\fh52  \\
2005-Jan-16 & 386.698 &     325    & 81     & 8\fh80  \\
2005-Jan-17 & 387.704 &     325    & 60     & 6\fh52  \\
\hline
\end{tabular}
\label{tab1}
\end{center}
\begin{tabular}{l}
\end{tabular}
\end{table}

   \begin{figure}
   \centering
   \includegraphics[width=8cm]{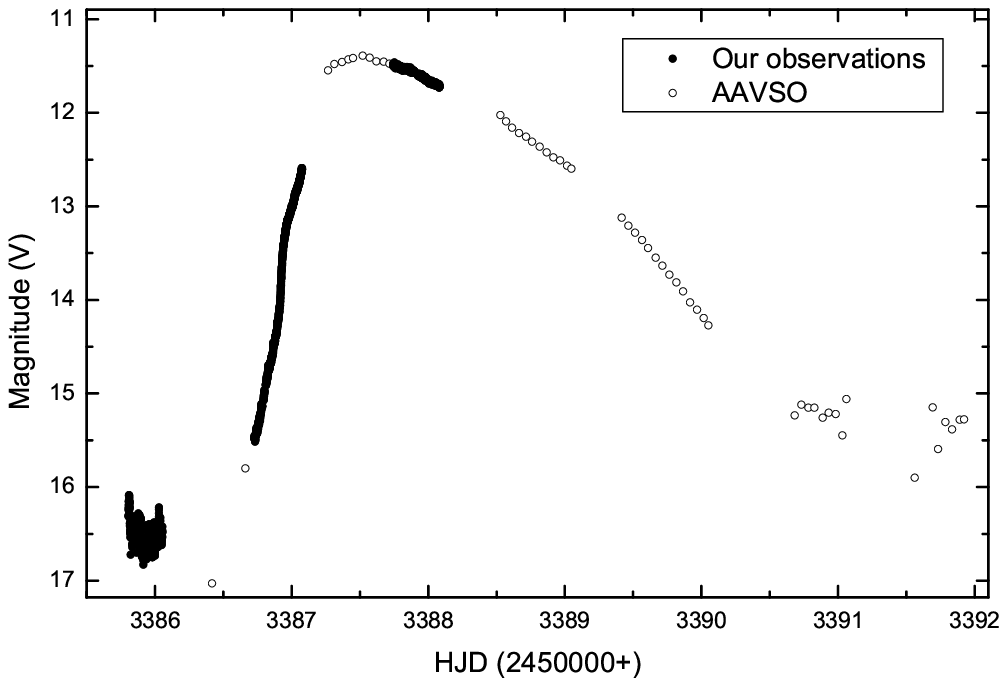}
      \caption{Light curve of BZ\,UMa during its Jan 2005 outburst.}
         \label{photometry}
   \includegraphics[width=8.0cm]{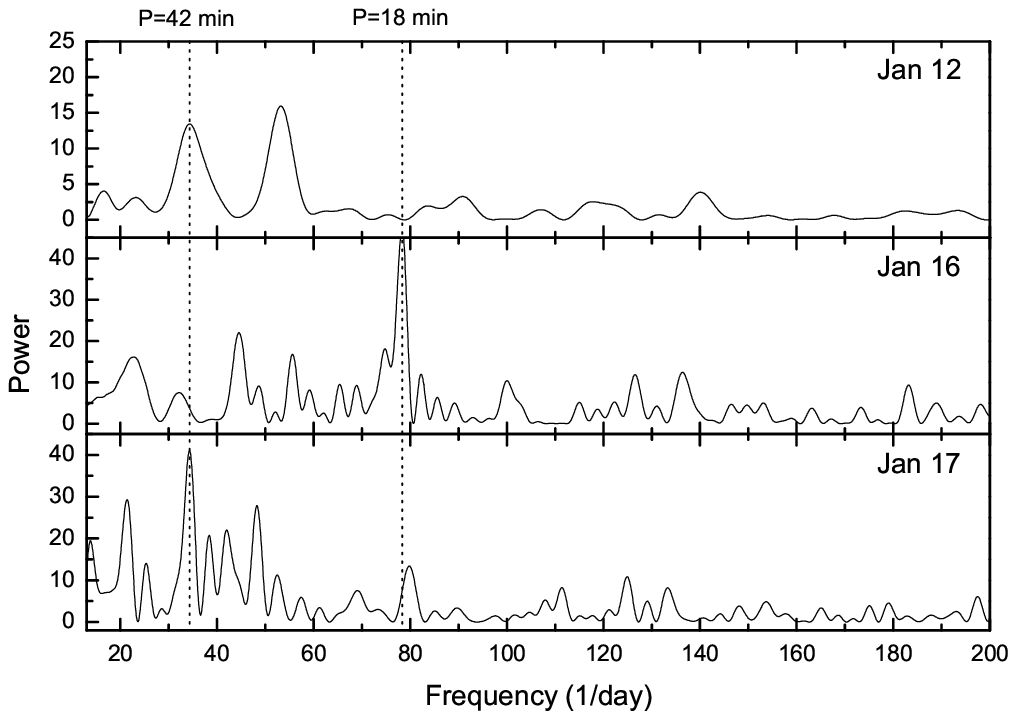}
      \caption{Power spectra of the nightly light curves.}
         \label{power}
    \includegraphics[width=8.0cm]{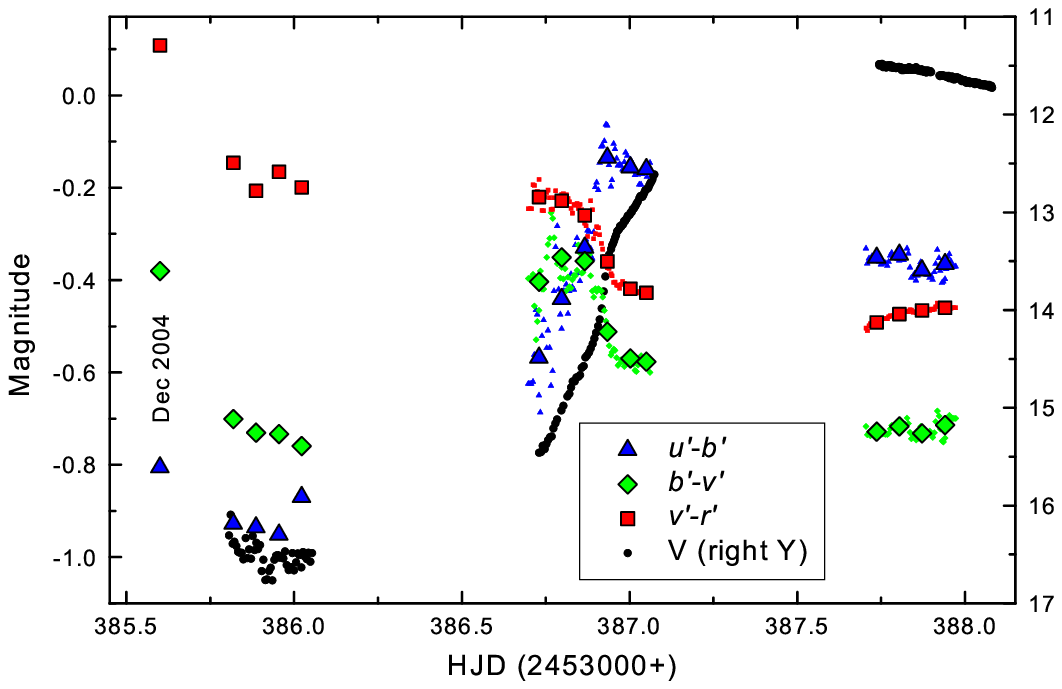}
      \caption{$u'-b'$, $b'-v'$ and $v'-r'$ light curves during the outburst.
       Large symbols represent data obtained from the \textit{period} averaged spectra
       while small ones are from the individual spectra.
       The V light curve is also shown for reference.}
         \label{colours}
   \end{figure}

    \begin{figure*}
    \centering
    \includegraphics[width=17.0cm]{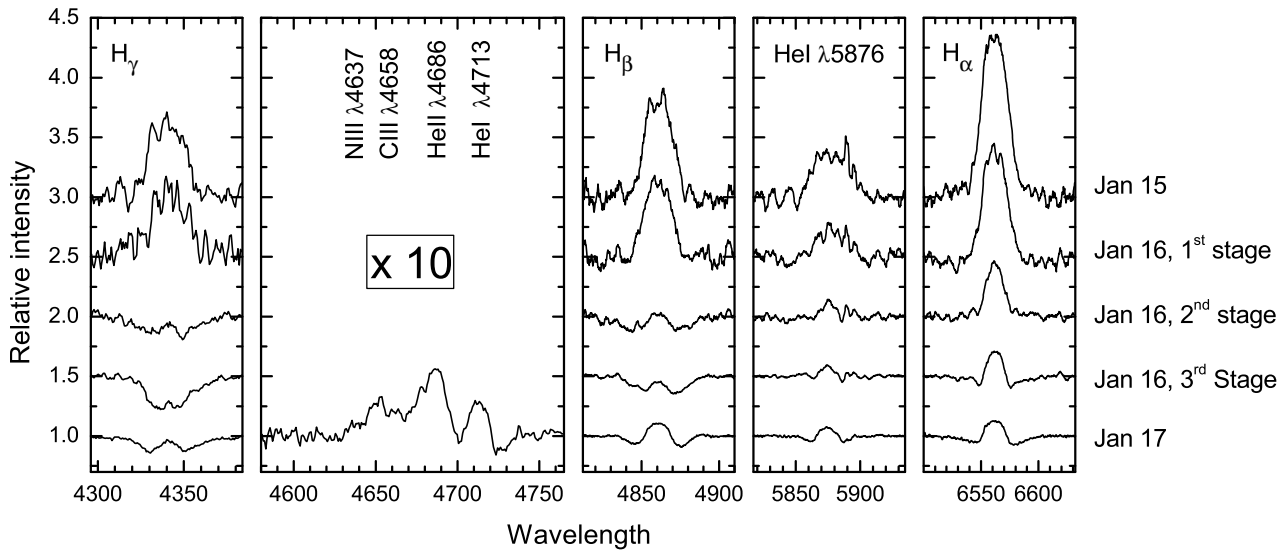}
      \caption{Averaged profiles of H$\gamma$, H$\beta$, He\,I $\lambda$5876, H$\alpha$ and
       area around the C\,III/N\,III -- He\,II -- He\,I $\lambda$4713 complex normalized
       to the continuum. The first spectra shown at the top are averaged for Jan 15,
       the following ones for the first stage of the outburst (three first periods of Jan 16),
       for the jump, the third outburst stage, and for Jan 17. The first spectra (Jan 15)
       are scaled down by a factor of 4 in relative intensity for display purpose and
       shifted upward by 2.0 units. Each subsequent spectrum is shifted upward by 0.5 units
       less. The C\,III/N\,III -- He\,II -- He\,I $\lambda$4713 complex are shown only for
       Jan 17, and are scaled up by a factor of 10 in relative intensity.}
      \label{Profiles}
   \end{figure*}

    The reduction procedure was performed using IRAF. Comparison spectra of Th-Ar
    lamps were used for the wavelength calibration.
    The absolute flux calibration of the spectra was achieved by taking nightly
    echellograms of the standard stars HD93521 and HD19445
    (actually, only HD93521 was used for calibration, while HD19445 was as a control star).
    Though we used a wide slit (2\arcsec) with seeing usually noticeably less than the slit
    width, this does not warrant excellent flux calibration, since only an average curve for
    atmospheric extinction and a permanent E-W orientation of the slit were used.
    At the same time, due to an unexpectedly observed outburst we found it useful to obtain
    some colour informations from our spectra. For this we defined an internal photometric
    system comprising four colour bands $u'\,b'\,v'\,r'$ centered at 3850\,\AA, 4550\,\AA,
    5590\,\AA, 6380\,{\AA} with widths of 50\,\AA, 100\,\AA, 120\,\AA, 150\,{\AA}
    respectively. The colour indices were calculated as $C=-2.5\log(f_1/f_2)$,
    where $f_1$ and $f_2$ are the fluxes averaged across the corresponding bands.
    To check the stability and the flux calibration accuracy we have determined
    the $u'-b'$, $b'-v'$ and $v'-r'$ colours of the control star HD19445 using
    our nightly spectra, and compared these with its published spectral energy
    distribution. The colours did not differ by more than 0.03 mag between any of our
    observations, and our colours were within 0.06 mag of the published spectra.

    To improve the confidence of the results presented in this paper we also
    acquired the \textit{period} averaged spectra obtained by means of co-adding of
    15 consecutive spectra. Additionally we obtained the phase-averaged spectra for
    the first three periods of Jan 16, which corresponds to the first stage of the
    outburst rise (see below).


\section{Results}

\subsection{Light curves and power spectra}

    During the first three nights of observations (Jan 12, 13 and 15) the system appears to
    have been in a quiescent state. Even on Jan 15, just before the outburst,
    the medium brightness remained almost as before: $\sim$16.5 mag.
    At the beginning of the next night (Jan 16) we found the system to be $\sim$1 mag
    brighter than 16 hours previously. During the following 8.5 hours the
    brightness increased by another 2.9 mag. This rising can be clearly divided into
    3 stages. In the first stage (during 4.3 hours) the flux was rising almost linearly at
    a rate of $\sim$0.34 mag hour$^{-1}$. This was followed by a `jump' when during 30-40
    minutes the brightness increased by 0.7 mag. Subsequently we observed a linear
    increase in luminosity but with a rate less than during the first stage:
    $\sim$0.2 mag hour$^{-1}$. However, according to amateur observations
    (Fig.~\ref{photometry}), the rate increased again just prior to maximum resulting
    in a maximum flux of $\sim$11.4 mag at about JD 2453387.52.
    We continued our observations $\sim$6 hours after the maximum and observed
    the decline with a nearly constant rate (in flux units).

    We calculated the power spectra for each detrended night's light curve and found 
    no evidence for orbital variability or superhumps.
    To the best of our knowledge, there is only two sets of BZ UMa's photometry
    during outbursts \citep{Kato, Price}. In both cases the authors reported an
    appearance of (quasi)-periodic oscillations (QPO) on the decline from outbursts
    with the period of oscillations close to 0\fd03 ($\sim$42 min).
    These oscillations were also observed by us during the decline,
    and also before the outburst (on 12$^{th}$ and 13$^{th}$ of January).
    Additionally, during the rise and probably after the maximum we have detected strong
    enough oscillations with the period of $\sim$18.4 min (Fig.~\ref{power}).
    We have found that
    \citet{Price} had also observed similar oscillations during
    the decline stage of the previous outburst though they did not mention
    the strongest peak in their power spectrum for unknown reasons (see Fig.5 in
    their paper).

    We have compared the light curves in all four spectral passbands and the
    $V$-band and found no time delays between them in any of the outburst stages.
    But the colour indices demonstrate dramatic changes with time
    (Fig.~\ref{colours}). One can see noticeable reddening of the flux distribution
    during the first outburst stage which stopped and even turned back to blueing just
    after the `jump'.

\subsection{Spectral changes}

    The spectrum of BZ\,UMa in quiescence is dominated by extremely strong hydrogen emission
    lines with a flat Balmer decrement. Apart from hydrogen, numerous fairly strong emission
    lines of neutral helium are present, and in addition to them also weak He\,II $\lambda4686$
    are observed \citep{Bruch}.

\begin{figure}
    \centering
    \includegraphics[width=8cm]{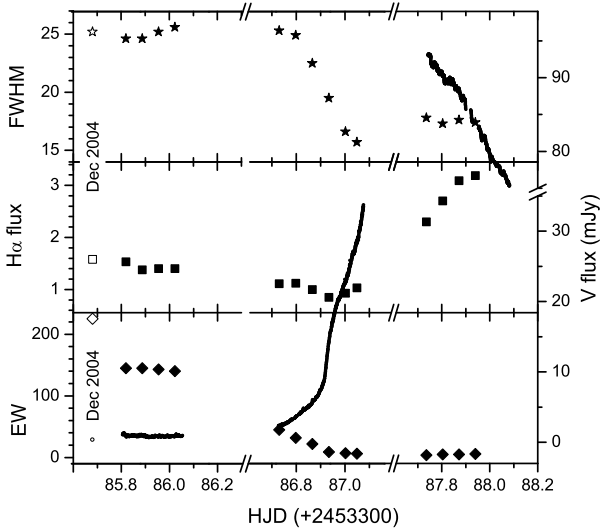}
      \caption{FWHM, line fluxes and EW of the H$\alpha$ emission during
      the 2005 January outburst (stars, squares and diamonds, respectively). The V light curve 
      (dots, right Y) is shown for reference. Open symbols represent December's values.}
      \label{line_param}
   \end{figure}

    Fig.~\ref{Profiles} shows the changes of the averaged profiles of the major
    spectral lines of BZ\,UMa during all the outburst stages. Prior to the outburst
    the spectrum looks like the usual spectrum for BZ\,UMa's quiescent state with
    strong Balmer and He\,I emission lines. During the first stage of the rise there was
    little qualitative change in the spectrum. In a quantitative sense these changes
    became apparent by the decreasing of FWHM, EW and flux of the emission lines
    (Fig.~\ref{line_param}).
    The qualitative changes appeared during and after the jump
    when the broad absorption troughs showed up around the Balmer and He~I emission lines.
    Their full width at continuum level corresponds to a velocity of ~3500 km s$^{-1}$.
    This is very close to the value for the wings of emission lines during quiescence
    and can be explained by broadening effects due to rapid rotation of particles in
    the innermost parts of the optically thick accretion disc. After the jump the line
    flux began to grow and during the following rise
    all emission lines with the exception of H$\alpha$ were present deep in the absorption
    troughs but never disappeared. However, the emission lines became stronger relative to
    the absorption lines after the outburst maximum.

    Of particular interest is tracing the changes of the high excitation lines
    (such as the He\,II and C\,III/N\,III blend emissions) which are good tracers of
    irradiation. Unfortunately, due to the weakness of these lines and the poor
    sensitivity of the CCD at short wavelengths we were able to detect their presence only during
    the decline stage of the outburst. However, having in hand the published low-resolution
    spectra of BZ\,UMa in a quiescent state \citep{Bruch, Jurcevic, Ringwald, Neustroev1}
    we can at least note that although absent in quiescence, the C\,III/N\,III Bowen blend has
    appeared during the outburst.
    But unlike some other dwarf novae which show strengthening of the He\,II
    and C\,III/N\,III line emissions during an outburst, these lines have remained quite
    weak in the outburst spectra of BZ\,UMA (Fig.~\ref{Profiles}).

    We also note that in spite of the high spectral resolution, all the emission
    lines show not the double-peaked profiles but are rather multi-peaked.
    However the profiles are highly variable.

   \begin{figure*}
   \centering
    \includegraphics[width=5.8cm]{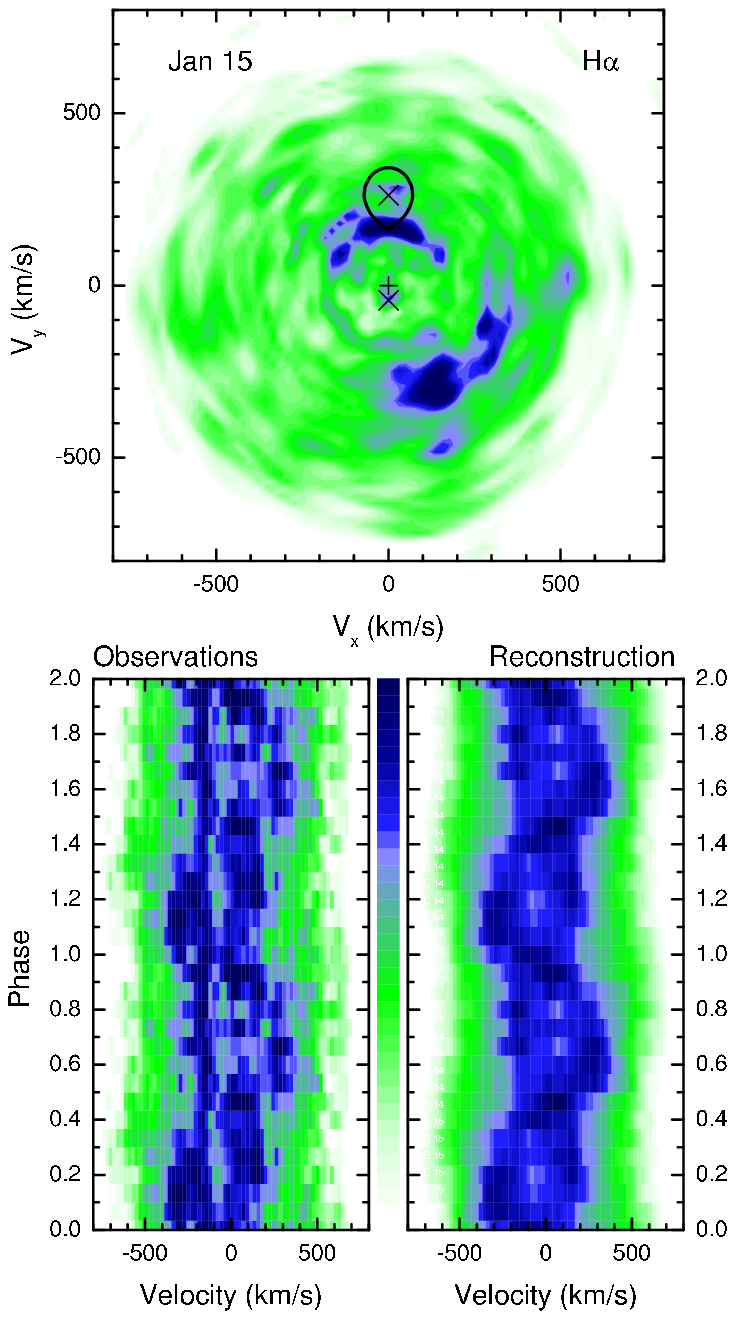}
    \includegraphics[width=5.8cm]{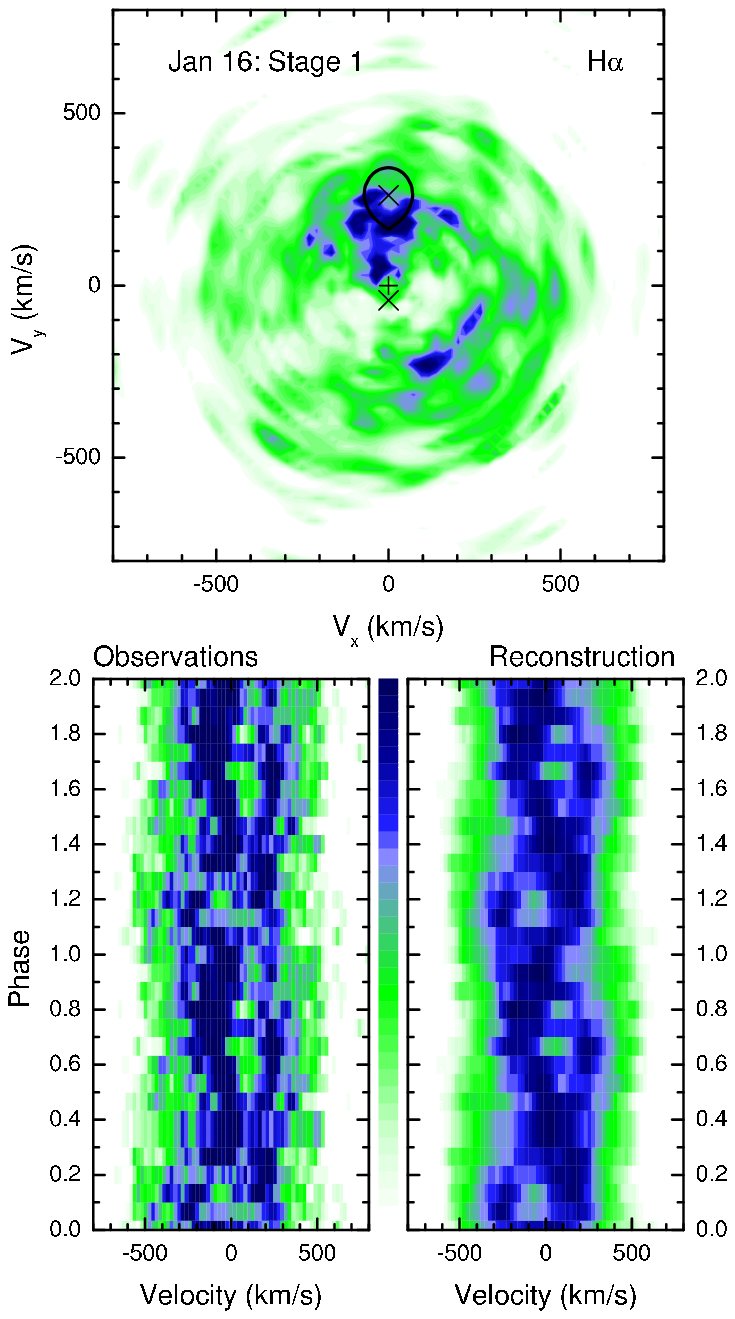}
    \includegraphics[width=5.8cm]{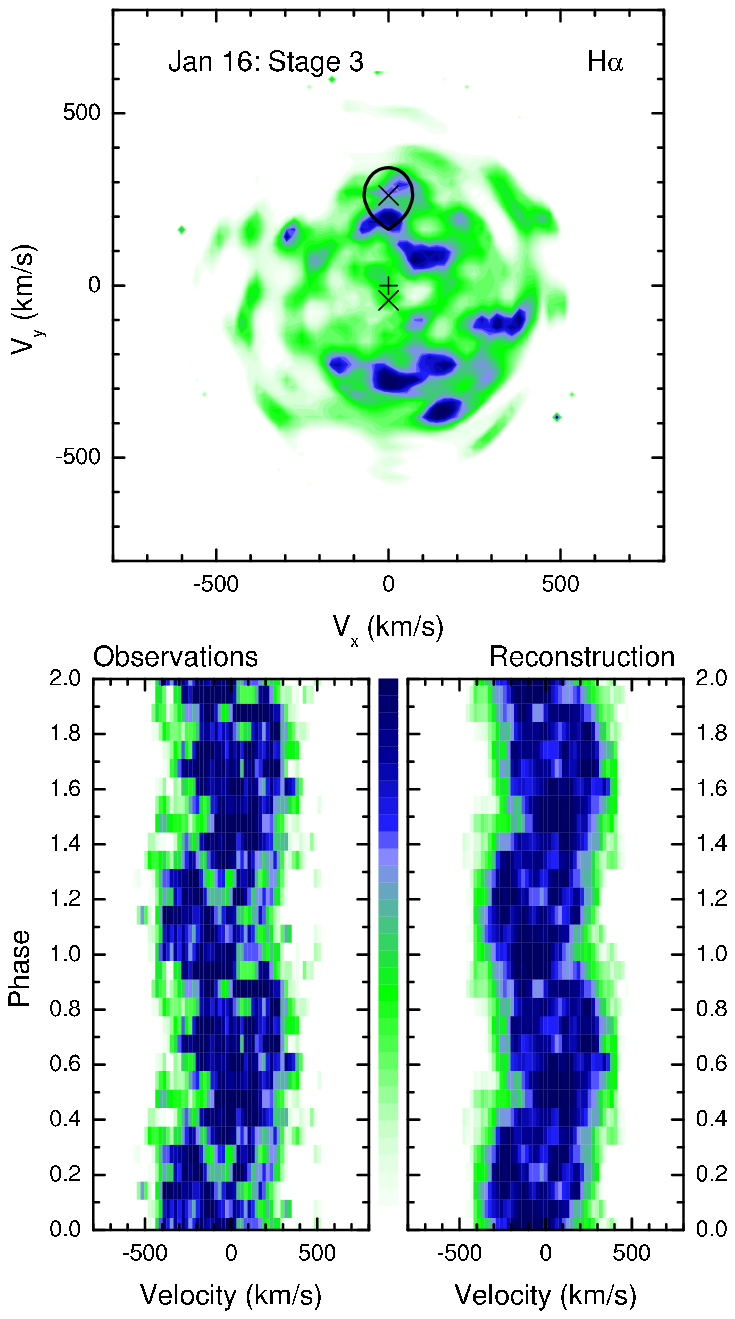}
    \caption{Observed and reconstructed trailed spectra (bottom)
    and corresponding Doppler maps (top) of BZ UMa before the outburst (Jan 15) and
    during the rise to it (Jan 16, stage 1 and 3).}
    \label{dopmaps1}
   \centering
    \includegraphics[width=5.8cm]{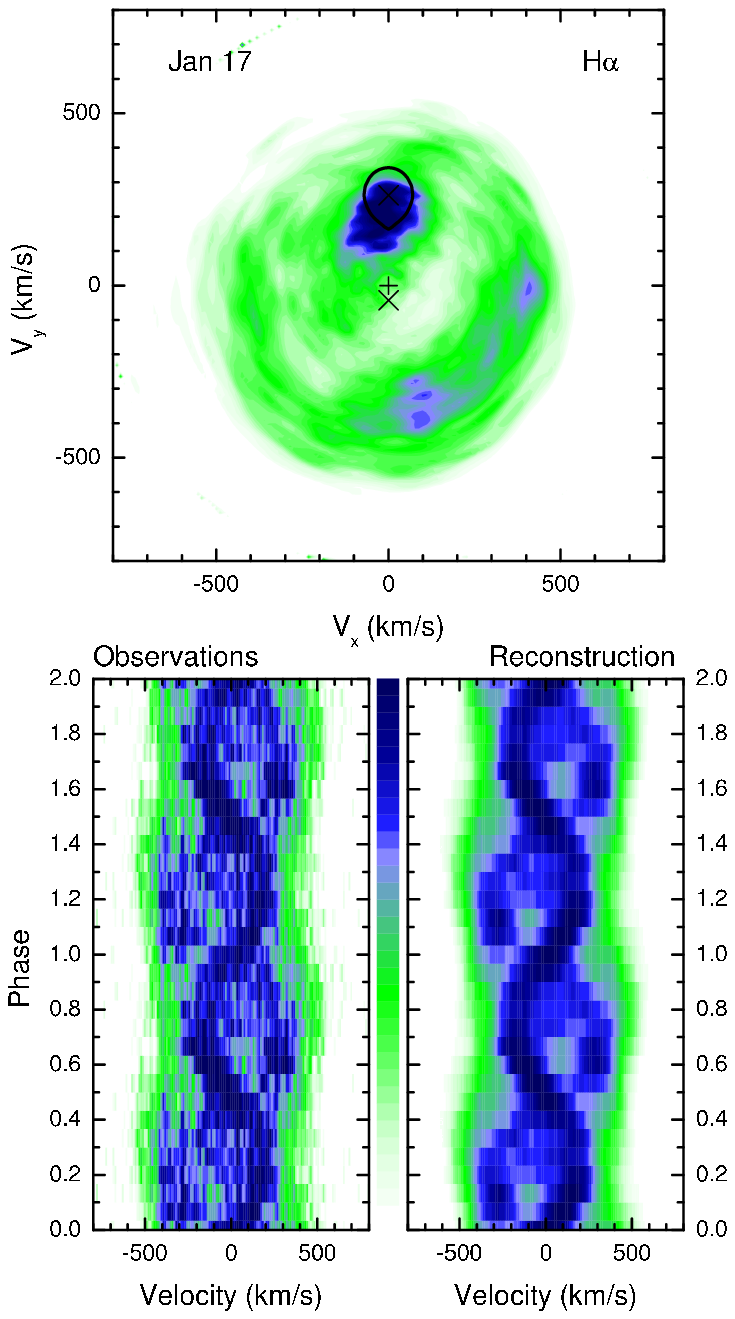}
    \includegraphics[width=5.8cm]{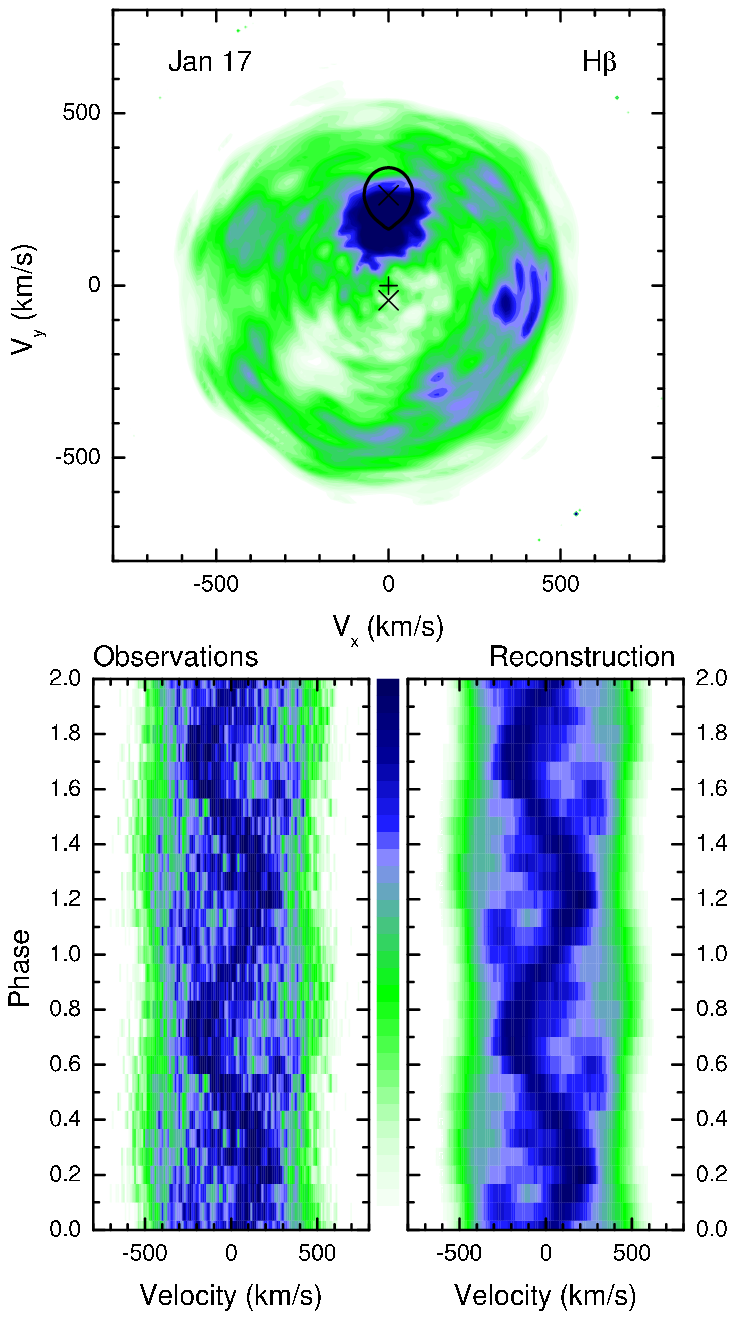}
    \includegraphics[width=5.8cm]{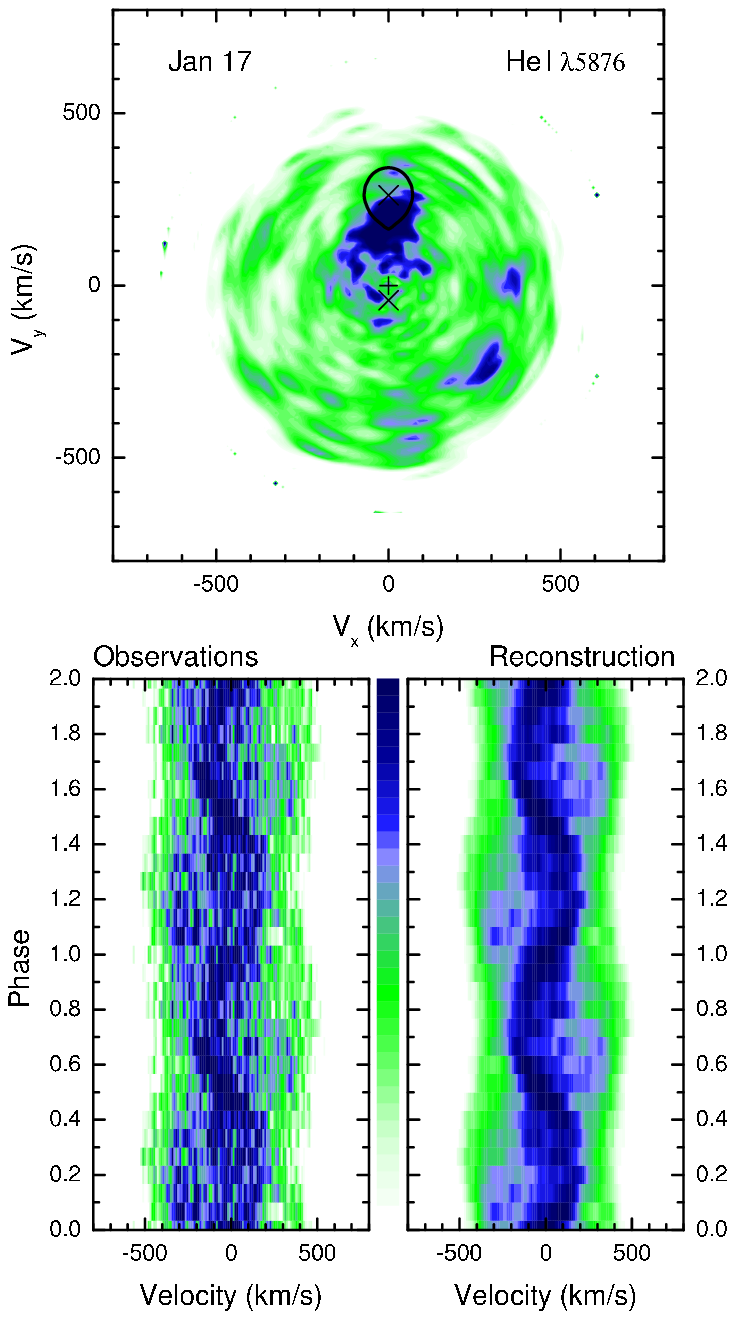}
    \caption{Observed and reconstructed trailed spectra (bottom) and corresponding
    Doppler maps (top) of BZ UMa after the maximum of the outburst (Jan 17).}
    \label{dopmaps2}

   \end{figure*}

 \subsection{Doppler Tomography}

    The orbital variation of the emission lines profiles indicates a non-uniform
    structure for the accretion disc.
    In order to study the emission structure of BZ\,UMa and its change during the outburst
    we have used Doppler tomography. Full technical details of the method are given by
    \citet{Marsh-Horne} and \citet{marsh2001}. Examples of the application of Doppler
    tomography to real data are given by \citet{marsh2001}.

    Figures~\ref{dopmaps1} and \ref{dopmaps2} show the tomograms computed using
    the code developed by \citet{Spruit}. These figures also show trailed
    spectra in phase space and their corresponding reconstructed counterparts.

    The preoutburst map (Fig.~\ref{dopmaps1}, left) displays a quite unusual and very
    nonuniform distribution of emission. Due to the non-double-peaked emission line profiles
    of BZ\,UMa, we did not expect a Doppler map to have an annulus of emission centered on
    the velocity of the white dwarf, and the observed tomogram does not show it.
    Instead of this the tomogram shows two extended bright areas.

    The first bright area, centered on (V$_x \approx 0$ km s$^{-1}$,
    V$_y \approx 170$ km s$^{-1}$), looks like a segment of a circle and occupies an area
    extending from azimuths about $315^{\circ}$ to $45^{\circ}$ (the corresponding phase
    of the intersection of the line-of-sight with this bright region is about $0.9 - 1.1$).
    The source of this feature is not clear. There can be speculations on its stream origin,
    but the shape and position of this feature raise a doubt.

    The brightest area is located in an unusual place, on the bottom-right part of the map,
    in exactly the same place as observed for BZ\,UMa before \citep{Neustroev2}.
    This extended bright area has a complex multi-spot structure.
    This is far from the region of interaction between the stream and the disc particles.
    None of the theories predict the presence of any bright spots here, which are
    connected with such an interaction.
    Additionally, we would like to note the existence of emission
    from the area around the WD.

    Surprisingly, this extremely unusual emission structure remains present in all major
    details during all the outburst stages. After the maximum a new emission source
    arose which became the brightest at that time (Fig.~\ref{dopmaps2}). It is situated
    close to the first bright area on the quiescent tomogram but most likely they are not
    directly linked. The new source can be unequivocally contributed to emission
    from the inner hemisphere of the secondary star. This emission seen in hydrogen and
    neutral helium lines, is likely caused by increased irradiation from the accretion
    regions during the outburst.

\begin{figure}
    \centering
    \includegraphics[width=4cm]{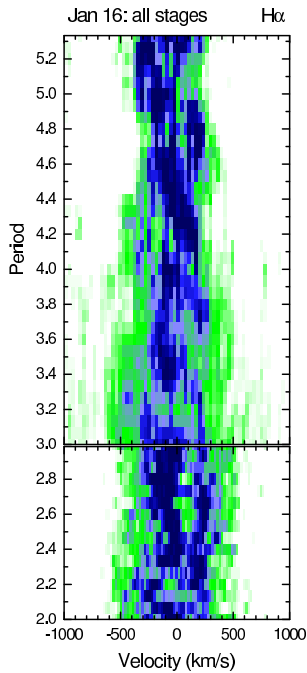}
      \caption{Observed trailed H$\alpha$ spectra of BZ UMa during the rise
      to the outburst (Jan 17). First three periods, averaged together into one period,
      are shown in the bottom panel, then all the other spectra are shown one after
      another in the upper panel.}
      \label{N2HaAll}
   \end{figure}

    It is interesting to trace what the emission from the donor star was doing during
    the rising stages of the outburst. Actually, it was incorrect to calculate the Doppler maps
    using these spectra because they have been obtained under conditions obviously breaking
    one of major principles of Doppler tomography: that flux from any point is constant in
    time \citep{marsh2001}. However, for display and comparison purposes we have produced
    the tomograms of the H$\alpha$ emission during the first and third stages of the outburst
    rising (Fig.~\ref{dopmaps1}, middle and right panels). They appear to be very similar
    to the quiescent map (Fig.~\ref{dopmaps1}, left), without any clear indications for the emission
    from the secondary. Figure~\ref{N2HaAll} separately shows all the trailed H$\alpha$ spectra
    from rise to outburst.

 \subsection{Radial velocities and zero phase}

    In CVs the most reliable parts of the emission line profile
    for deriving the radial velocity curve are the extreme wings. They
    are presumably formed in the inner parts of the accretion disc and therefore
    should represent the motion of the white dwarf with the highest reliability.
    We measured the radial velocities using the double-Gaussian method
    described by \citet{sch:young} and later refined by \citet{Shafter}.
    In order to test for consistency in the derived velocities and the zero phase,
    we separately used the emission line H$\alpha$ in December's spectra, in the
    phase-averaged spectra for the nights of Jan 15 and 17, and in the phase-averaged
    spectra for the first three periods of Jan 16.
    All measurements were made using a Gaussian FWHM of 100 ${\rm km~s^{-1}}$ and
    different values of the Gaussian separation $\Delta$ ranging from 300
    ${\rm km~s^{-1}}$ to 3000 ${\rm km~s^{-1}}$ in steps of 50 ${\rm km~s^{-1}}$,
    following the technique of `diagnostic diagrams' \citep{Shafter2}.
    For each value of $\Delta$ we made a non-linear least-squares fit of the derived
    velocities to sinusoids of the form $\gamma-K\sin\left[2\pi(T-T_0)/P\right]$
    with the orbital period fixed to 97.9 min, and we find the maximum useful
    separation to be $\Delta \simeq 1200{-}1400$ ${\rm km~s^{-1}}$.
    We obtain very consistent results for both the radial velocity
    semi-amplitudes and the $\gamma$-velocities for the quiescent and outburst
    states, and we adopt the mean values $K=43\pm2 {\rm km~s^{-1}}$ and
    $\gamma=-29\pm3 {\rm km~s^{-1}}$.

    In the previous Section we have detected the extremely unusual emission structure of BZ\,UMa
    (particularly, the spot(s) in the lower-right of the Doppler maps), so we must be sure
    of the correct phasing of the input spectra used for producing the tomograms.
    The best way to determine the zero phase is the eclipse observations but eclipses
    are not present in BZ UMa.
    Assuming that the wings of the emission lines come from disc material orbiting close
    to the white dwarf, the red-to-blue crossing of the radial velocities provides
    an estimate of the moment of inferior conjunction of the secondary star.
    In general, a value for the zero phase obtained in this way may be influenced
    by additional emission sources, if they exist on the line wings. However, we believe we
    could avoid this as the detected strong emission spot is situated well inside of the
    chosen Gaussian separation. Moreover, for the Doppler mapping we have phased
    the input spectra separately for each night, in accordance with their respective moments
    of inferior conjunction of the secondary star $T_0$, however all these $T_0$'s are
    consistent with each other and with the chosen orbital period of 97.9 min.
    Finaly we note that the correctness of the used zero phases are strongly supported by
    the detected emission from the donor star.

\begin{table}
\caption[]{Elements of the radial velocity curves of BZ UMa}
\begin{flushleft}
\begin{tabular}{cccc}
\hline
\noalign{\smallskip}
Date & $\gamma$-velocity & $K$ & T$_{0}$ \\
     & (km s$^{-1}$) & (km s$^{-1}$) & (HJD) \\

\noalign{\smallskip}
\hline
\noalign{\smallskip}
2004-Dec-10 & -42$\pm$8 & 38$\pm$9 & 2450101.264$\pm$0.001 \\
2005-Jan-15 & -27$\pm$5 & 42$\pm$6 & 2450101.265$\pm$0.001 \\
2005-Jan-16 & -27$\pm$4 & 42$\pm$7 & 2450101.263$\pm$0.002 \\
2005-Jan-17 & -55$\pm$2 & 44$\pm$3 & 2450101.265$\pm$0.004 \\
\noalign{\smallskip}
\hline
\noalign{\smallskip}
\textbf{Mean} & \textbf{-29$\pm$3} & \textbf{42.8$\pm$3} & \textbf{2450101.264$\pm$0.001} \\

\noalign{\smallskip}
\hline

\end{tabular}
\label{Table3}
\end{flushleft}
\end{table}

\section{Discussion}

    The nonperiodic nature of outbursts makes any observations of 
    their early stages very important. The presented simultaneous photometric 
    and high resolution spectroscopic observations should be useful to constrain 
    current models of outburst behaviour.

    Although significant theoretical advances have been made in 
    recent years, there remains many unsolved problems
    (see review of \citealt{Lasota}). One of the major issues is that
    the system brightness in the models almost always increases between the 
    outbursts but the optical observations show that the mean flux remains 
    approximately constant throughout quiescence. In the case of BZ\,UMa,
    the changes in the system had already begun before the outburst event.
    On the basis of the
    December spectroscopy we have determined that the system was redder and
    $\sim$0.9 mag fainter one month before the outburst than in the last
    few `quiescent' days preceding the outburst.
    Moreover, though the hydrogen emission lines prior to the outburst still remained
    extremely strong (EW$_{H\alpha}\sim140$\,\AA), they were the weakest we have ever
    observed \citep{Neustroev1, Neustroev2}. For example, in December the EW$_{H\alpha}$ 
    was $\sim$225\,\AA. 
    These changes were not caused by an enhancement of the mass-transfer rate:
    The Mass-Transfer Instability model predicts an increase of the brightness 
    of the hot spot before and during the outburst but the Doppler maps 
    (Fig.~\ref{dopmaps1} and \ref{dopmaps2}) do not show any sign of
    the hot spot. At the same time, in the Disc Instability Model no such effects 
    are expected.
    
    According to the DIM, there can be two types of outbursts
    depending on where the instability in the accretion disc begins.
    When the trigger is in the inner disc, a heating wave pushes outward,
    creating an `inside-out' outburst; if the trigger is in the outside
    of the disc, the heating wave pushes inward, creating an `outside-in'
    outburst. Which type of outbursts occur depends sensitively on the mass
    transfer rate from the secondary star and the viscosity. An outside-in
    outburst results if the transfer rate is high while the inside-out outburst
    occurs if the viscous evolution within the disc is more effective. For
    constraining these parameters, it is important to be able to determine
    which of these types of observed outbursts is occurring in a particular
    system.

    The large outburst amplitude and the long recurrence time of BZ\,UMa, 
    together with one of the strongest Balmer emission spectra of any known CV, 
    indicates a low accretion rate for the system \citep{Patterson}.
    As $\dot{M}$ is an all-important parameter in the following discussion, 
    we present our considerations concerning its value in BZ\,UMa.      
      
    In order to estimate $\dot{M}$ one needs to know the distance to the system.     
    Distances can be derived from the absolute magnitude at outburst
    $M_V$(max) versus the orbital period relation of \citet{Warner} or from 
    a more recent relationship by \citet{Harrison}.
    Analysis of the last sixteen outbursts of BZ\,UMa, particularly documented in
    the AAVSO database, reveals a roughly uniform distribution
    of peak magnitude from 10.5 mag to 11.5 mag.
    This yields 110--175 pc.\footnote{We suppose that smaller values are more probable
    if outbursts of lower magnitude correspond to the case when the heating front
    does not reach the outer disc regions (see discussion below).}
    These independent estimations are consistent with the $110^{+44}_{-51}$ pc value
    of \citet{Ringwald}, leading us to the final estimation of 110--154 pc.
    To determine $\dot{M}$ we could use a standard-candle method.
    The dwarf nova HT\,Cas shows many close similarities to BZ\,UMa including
    the outburst behaviour and system parameters, thus their comparison could be useful
    to determine $\dot{M}$. The mass-transfer rate in HT\,Cas is
    $\sim2\times10^{15}$ g s$^{-1}$ \citep{Wood} 
    but this system shows the outburst amplitude and
    the EW of emission lines substantially less than in BZ\,UMa, and it appears to
    be more intrinsically bright than BZ\,UMa. From this we expect $\dot{M}$ for BZ\,UMa
    to be several times less than in HT\,Cas. 
    Using the approximate method of
    \citet{Patterson} we derive $\dot{M}$ for BZ\,UMa to be even less than
    10$^{14}$ g s$^{-1}$.
    
    A very low mass-transfer rate for BZ\,UMa implies that theoretically, 
    its outbursts should be of an inside-out type.
    Unfortunately, the only way to be sure of the outburst type is from eclipse
    profile observations during outbursts and BZ UMa is not an eclipsed system.
    Nevertheless, there is some additional circumstantial evidence to support
    the inside-out type of the outburst of BZ\,UMa.
    
    The strongest indication comes from the spectra obtained during the rising stage.
    For the case of the inside-out outburst one can expect the lines wings change
    before their cores. Such is indeed the case in BZ\,UMa (Fig.~\ref{Profiles}). 
    Already during the jump the wings of the emission lines were replaced by the
    broad absorption troughs. This can be explained by the transition 
    of the innermost parts of the accretion disc from a cool, optically thin 
    state to a much hotter and optically thick state. This transition on the earliest 
    stages of an outburst can occur only in the case of the inside-out type.
    
    In favour of the inside-out outburst may speak also a noticeable difference
    in peak magnitudes of BZ UMa during outbursts (see above). \citet{Smak2}
    showed that there exit two types of the inside-out outbursts: in one
    the heating front does not reach the outer disc regions while in the other
    the front propagates all the way to the outer edge. Consequently, the
    first type is of lower magnitude and shorter duration than the second,
    conforming with the observations.   
  
    However, some of our observations raise an interesting and related question.
    Theory predicts that the inside-out outburst produces a fairly symmetric
    light curve with respect to rise and decline while the outside-in outburst
    develops more quickly and produces an asymmetric light curve \citep{Smak2}.
    Actually, the outburst light curve of BZ UMa is highly asymmetrical,
    with the rise of $\sim0.8$ days and the decline of $\ga3.5$ days, relative
    to a level of 15.5 mag, testifying against the inside-out type.
    In principle, this guess may be correct because the outside-in outburst has
    been directly observed in HT\,Cas \citep{Ioannou}, another system with a
    low mass-transfer rate. Such an outburst in this system can be explained only    
    if the disc is significantly smaller than the tidal truncation radius \citep{Buat}.
    This is in agreement with observations of HT\,Cas but it is unlikely for BZ\,UMa.
    BZ\,UMa exhibits a transfer rate well below the level of the transition
    between inside-out and outside-in outburst, and we do not see how one could get
    outside-in outbursts in this system even with the small accretion disc.

    It is known that in some conditions inside-out outbursts may produce
    an asymmetrical light curve. This is possible when the inner disc
    is truncated, for example due to the presence of a magnetic field
    \citep{Lasota2}. We found no clear sign of the truncated accretion disc, though
    evidence for a weak magnetic WD in BZ UMa may be
    indicated by its relatively strong X-ray emission \citep{Verbunt},
    by the presence of modulations with the period of 42 min (and/or 18 min),
    and by emission from an area around WD in the Doppler maps.

\section{Summary}

    In this paper we reported simultaneous photometric and echelle-spectroscopic 
    observations of the dwarf nova BZ UMa during which we were lucky to catch 
    the system at the onset of an outburst. On the base of the spectral changes 
    during the rising stage we conclude that the 2005 January outburst of BZ UMa 
    was of the inside-out type. Nevertheless, a highly asymmetrical light curve 
    with a `jump' during which the system brightness almost doubled over the 
    course of half an hour, shows that this outburst cannot easily be fit into 
    the framework of the current Disc Instability Model.

\section*{Acknowledgments}

    VN acknowledges support of IRCSET under their basic research programme and
    the support of the HEA funded CosmoGrid project.
    We acknowledge with thanks the variable star observations from the AAVSO
    International Database contributed by observers worldwide and used in this
    research. The authors would like to thank Jean-Pierre Lasota and the referee 
    for their useful comments, and Padraig O'Connor for improving the language of 
    the manuscript. We are grateful Felipe Montalvo for the assistance during 
    the observations.

\bsp
\label{lastpage}
\end{document}